# RACiMo: **R**ed **A**mbiental **Ci**udadana de **Mo**nitoreo:

## A Student-Centred Citizen Science Network for Environmental Monitoring, Data Literacy, and Climate Awareness in Colombia

**L.V. Flórez, G. Sánchez-Ariza, N. Mantilla-Molina, J. Pisco-Guabave,**
**A. Martínez-Méndez, A. Serrano-Mendoza, Luis A. Núñez**[1]
Escuela de Física, Universidad Industrial de Santander, Bucaramanga, Colombia.
**H. Asorey**
piensas.xyz & Fractalis Adventures Las Rozas Innova, Las Rozas de Madrid, Madrid, Spain
**L.M. Becerra**
Centro Multidisciplinario de Física, Universidad Mayor, Santiago, Chile
**J. Grisales-Casadiegos**
Departamento de Astronomía, Universidad de Guanajuato, Guanajuato, México.
**A. Núñez-Castiñeyra**
Institut d'Astrophysique de Paris, Sorbonne Université, 75014 Paris, France
**J. Peña-Rodríguez**
Faculty of Physics, University of Wuppertal, Gaußtraße 20, 42119 Wuppertal, Germany,
**J. Salamanca-Coy**
Sentidos Innovadores e Inteligentes SAS, Bucaramanga, Colombia
**C. Sarmiento-Cano**
Departamento de Ciencias Básicas, Universidad Autónoma de Bucaramanga, Colombia
**D. Sierra-Porta**
Departamento de Ciencias Básicas, Universidad Tecnológica de Bolívar, Cartagena, Colombia

**Abstract:** Climate change and air pollution increasingly require local capacities for environmental monitoring, data interpretation, and community-based adaptation. This paper traces the evolution of Red Ambiental Ciudadana de Monitoreo (RACiMo), a citizen-science and environmental-education initiative developed by the Halley Research Group at the Universidad Industrial de Santander in Colombia. Emerging from early climate-awareness activities in 2013–2014, RACiMo has evolved through successive editions from a school-based open-hardware programme into a regional environmental monitoring network that integrates meteorological and air-quality data, open-data practices, and student-centred research. The project's central objective is to train citizens, especially secondary-school students, teachers, and university mentors, to record, curate, analyse, interpret, and communicate climate and air-quality data relevant to their local contexts. Across its trajectory, RACiMo has involved approximately 515 students from urban, metropolitan, rural, and páramo communities in Santander, Colombia. Its technological evolution has progressed from Arduino–Raspberry Pi stations assembled through a Do-It-With-Others approach to low-maintenance commercial instruments and a multi-municipality network of professional weather and air-quality stations. In parallel, its pedagogy has shifted from learning by building sensors to learning by analysing real environmental datasets through Python, Jupyter notebooks, visualisation tools, and student-led research projects. The current RACiMo-Orquídeas implementation expands the network across five municipalities, emphasises local environmental issues such as freight traffic, and industrial activity, and provides public access to data through repositories and interactive tools. The paper discusses the trade-offs between openness, educational value, reliability, data quality, sustainability, and scalability. It argues that citizen environmental monitoring can strengthen climate awareness and adaptation when local communities are trained not only to collect data, but also to understand and use it critically.

[1]Escuela de Física, Universidad Industrial de Santander, Bucaramanga, Colombia. Corresponding author e-mail: lnunez@uis.edu.co

# 1. Introduction

Climate change is altering the environmental conditions under which urban, peri-urban, and rural communities live. Its effects are expressed not only through changes in temperature, precipitation, and extreme events but also through interactions with air quality, public health, water availability, agriculture, and local livelihoods. Poor air quality, in particular, can amplify the risks associated with global warming and has become a central concern in public-health and sustainability agendas (Tran et al., 2023). In this context, preparing young people and local communities to understand environmental change, interpret local evidence, and participate in adaptation processes is an urgent educational and social challenge (Sezen-Barrie et al., 2025).

Environmental monitoring has traditionally been the responsibility of official institutions that operate national or regional sensor networks. In Colombia, this role is fulfilled by the Instituto de Hidrología, Meteorología y Estudios Ambientales (IDEAM), which also maintains a public data policy through national open data platforms. These institutional networks are indispensable for climate and environmental governance. However, their spatial and temporal coverage is often insufficient to characterise local-scale phenomena, particularly in mid-sized cities, peri-urban areas, smaller municipalities, and rural territories where environmental pressures may be highly heterogeneous (Muller et al., 2015; Kumar et al., 2015; Costa et al., 2015). In such contexts, complementary monitoring strategies based on low-cost sensors, distributed stations, and local participation can help improve environmental awareness and data availability, specially in the global south (Rojas González & Montilla-Rosero, 2025; Morresi et al., 2025; Tarazona-Alvarado et al., 2024; Sierra-Porta et al., 2023).

The usefulness of environmental data depends not only on the existence of sensors, but also on the capacity of communities to understand how data are produced, curated, visualised, interpreted, and communicated. This requirement is particularly relevant for climate adaptation, where local decisions often depend on the ability to connect measurements with everyday experience: heat, rainfall, particulate matter, traffic, industrial activity, agricultural conditions, or ecosystem vulnerability. Citizen science provides a framework for this process by engaging non-academic participants in generating knowledge relevant to both science and society (Prandi et al., 2025). When properly supported, citizen science can expand data-collection capacity, strengthen scientific literacy, and democratise access to environmental information (Prandi et al., 2025; Soacha-Godoy et al., 2025).

Over the last two decades, people-centric and human-powered sensing approaches, including crowd-sourcing, participatory sensing, crowd-sensing, and collective intelligence, have shown that citizens can contribute to the production, verification, analysis, and dissemination of environmental data (Campbell et al., 2008; D'Hondt et al., 2013; Guo et al., 2015; Delmastro et al., 2016). These approaches have also motivated the development of citizen observatories, FabLab-based initiatives, academic spin-offs, and civic monitoring platforms in several urban contexts (Mominó et al., 2016; Lanfranchi et al., 2014; Liu et al., 2014; Wehn & Evers, 2014;

Palacin-Silva et al., 2016). Examples from Amsterdam, Barcelona, Birmingham, Cambridge, the Ferrara region, and Singapore illustrate the diversity of institutional and community-based models through which environmental data can be generated and shared (Jiang et al., 2016; Diez, 2012; Chapman et al., 2015; Mead et al., 2013; Gravagnuolo et al., 2014; Kloeckl et al., 2011).

RACiMo, the Red Ambiental Ciudadana de Monitoreo (Citizen Environmental Monitoring Network) was conceived within this framework. Its central objective is to train citizens, especially secondary-school students, teachers, and university mentors, to participate in the full environmental data cycle: recording meteorological and air-quality variables, curating and organising the resulting datasets, analysing them using open computational tools, and interpreting them in relation to local environmental problems. Rather than treating communities only as passive recipients of environmental information, RACiMo seeks to create learning environments in which participants can understand where data come from, what their limitations are, and how to use them to support environmental awareness and adaptation to climate change.

This paper presents the evolution of RACiMo as a long-term citizen-science and environmental-education initiative developed by the Halley Research Group at the Universidad Industrial de Santander, Colombia. The project emerged from early activities in climate awareness and open data science and has evolved through successive editions in urban schools, rural páramo communities, and municipalities across the department of Santander in Colombia. Across this trajectory, RACiMo has combined three coupled processes: the technological evolution from community-built open-hardware stations to more robust professional monitoring devices; the pedagogical evolution from learning by building sensors to learning by analysing real environmental data; and the territorial evolution from school-based activities in Bucaramanga to a broader departmental network involving urban and rural communities.

The original contribution of this work is not limited to the description of a monitoring network. It is the analysis of RACiMo as a sustained model for building local environmental data capacity through citizen science. We examine how different technological choices affected community participation and data reliability; how training strategies were adapted to different educational and territorial contexts; how student-driven projects used real network data to formulate and communicate local environmental questions; and how open-data resources can support broader access to environmental information. In doing so, the paper argues that citizen environmental monitoring can contribute to climate adaptation when it is designed not only to collect data, but also to train communities to understand, question, and use those data.

## 2. Methodology

RACiMo was designed as a distributed environmental monitoring network in which the technological platform also functions as an educational instrument. From its first implementation, the project combined three operational components: (i) the deployment of local stations for recording meteorological and environmental variables, (ii) the preservation and dissemination of data through open or accessible repositories, and (iii) the training of students and teachers in the construction, configuration, calibration, interpretation, and use of the measurements. The first RACiMo design described the station as a modular system capable of operating with different levels of autonomy, so that each community could adapt the measuring and communication modules to its own interests, resources, and technical capacity (Asorey et al., 2017a; Peña-Rodríguez et al., 2022).

This design principle has remained constant throughout the project, although the hardware has changed substantially. The technological trajectory of RACiMo can be summarised as a transition from community-built, open-hardware stations to increasingly robust, low-maintenance commercial instruments. This transition was not merely technical. It modified the type of citizen participation promoted in each edition: in the earliest versions, students were directly involved in assembling and configuring the stations; in later versions, the educational focus progressively shifted toward station operation, data access, visualisation, and interpretation (Figure 1, Suppl. Table 1).

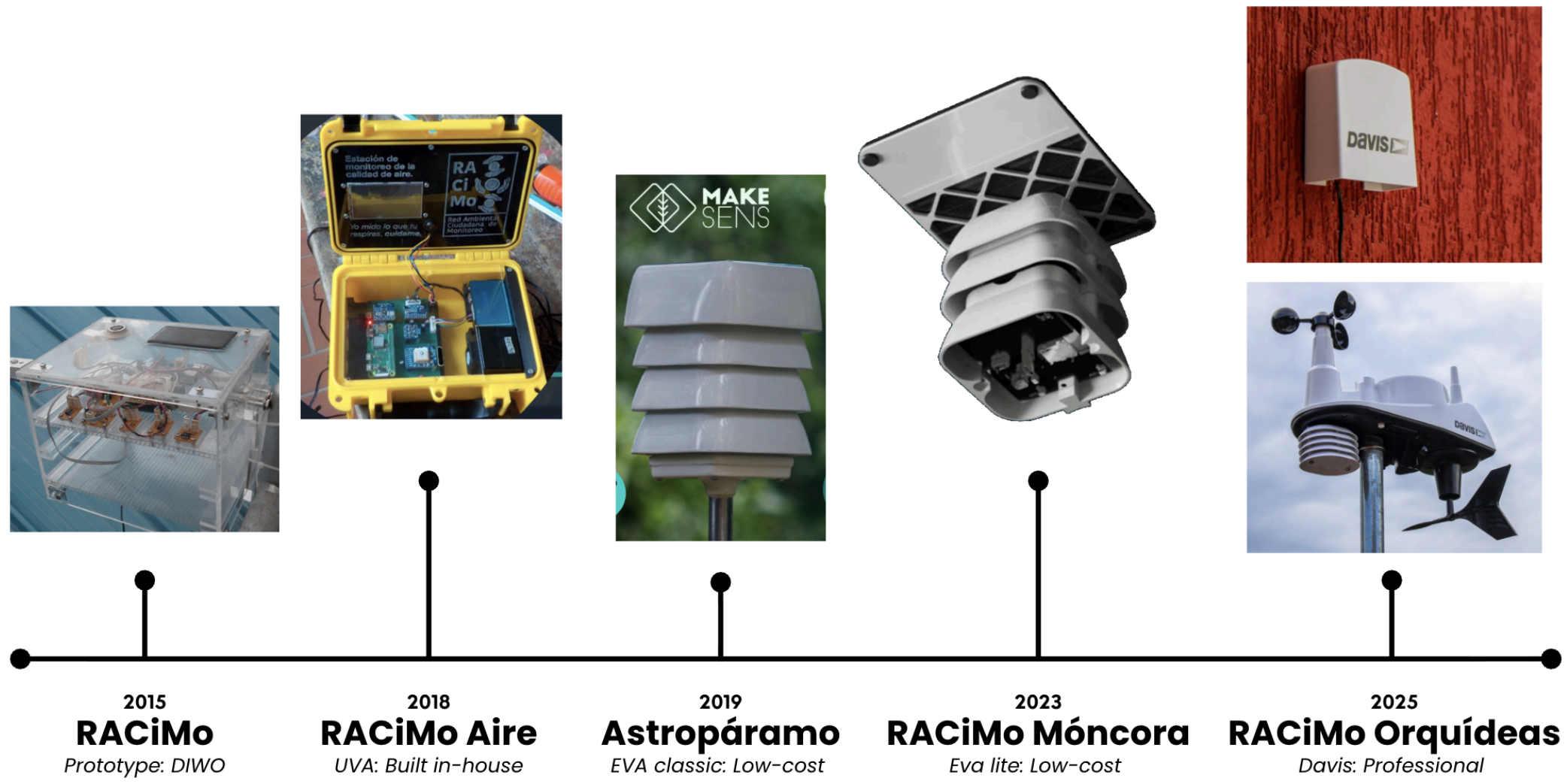

**Figure 1.** Timeline of the RACiMo project, showing the evolution of its monitoring devices across the five editions. The DIWO Prototype (RACiMo, 2015), the in-house-built UVA (RACiMo-Aire, 2018), the low-cost EVA Classic (Astropáramo, 2019) and EVA Lite (RACiMo-Móncora, 2023), both by MakeSens, and the professional Davis Instruments stations (RACiMo-Orquídeas, 2025).

## 2.1 Technological sequence of RACiMo editions

The first RACiMo stations were developed under a Do-It-With-Others (DIWO) approach. They were conceived as low-cost modular prototypes assembled with students and supervised by university mentors. The basic architecture combined an Arduino Uno board for sensor acquisition and pre-processing with a Raspberry Pi for data management, local storage, and transmission. The station included GPS, WiFi connectivity, a peripheral sensor shield, and a waterproof enclosure. In this first configuration, the main measured variables were atmospheric pressure and temperature (using an MPL115A2 sensor) and illuminance (using a light-dependent resistor). The station transmitted data through WiFi to a central repository, where the records could be accessed and analysed using Python-based tools (Asorey et al., 2017a; Peña-Rodríguez et al., 2022).

The second stage was the development of the UVA station, or Unidad de Valoración Ambiental. The UVA preserved the modular logic of the original RACiMo design but was built by the research team rather than assembled entirely by students. Its architecture continued to rely on two main blocks: an Arduino-based sensing block and a Raspberry Pi-based data-management block. The Arduino received signals from the sensors through standard communication protocols, including I2C, UART, analogue input, and digital input/output. The Raspberry Pi operated as an embedded Linux single-board computer responsible for storing, organising, and transmitting the acquired data. The use of SD storage allowed local data buffering, while network connectivity enabled data upload to external repositories (Peña-Rodríguez et al., 2022).

In the UVA configuration, the number and type of sensors increased. The station was designed to measure atmospheric pressure, temperature, relative humidity, precipitation, illuminance, irradiance, and cloudiness. Some sensors were commercial devices with direct calibration protocols, whereas others were in-house implementations. The irradiance sensor, for instance, used a small solar cell, while the cloudiness sensor used a Raspberry Pi camera and image-processing algorithms to estimate cloud cover from sky images. The data flow was automatic: sensor records were assembled into data packets, complemented with metadata, stored locally, and periodically transmitted. Metadata included station name, firmware version, location, connected sensors, power supply, GPS information, time zone, and data structure. Missing sensors or disconnected channels were marked with a predefined null value to preserve the data-file structure. A watchdog system checked the recording process, wireless connection, and data-upload routines, improving operational autonomy (Peña-Rodríguez et al., 2022).

RACiMo-Aire marked a further technological step by shifting the monitoring emphasis from meteorological variables toward air quality. This edition used an upgraded version of the UVA concept, capable of recording particulate matter, especially PM2.5 and PM10, together with temperature, relative humidity, and atmospheric pressure. It also incorporated a daily Air Quality Index. In this stage, data transmission moved toward a centralised hub accessed through web applications, with records exportable in formats such as JSON, CSV, KML, and GeoJSON. This transition improved data availability and enabled broader use of the records beyond the local station; it also anticipated later work on calibration and monitoring challenges associated with portable, modular and low-cost air-quality stations (Tarazona-Alvarado et al., 2024).

The Astropáramo edition introduced a different operational context. The project was implemented in rural schools in the Santurbán páramo region, where internet connectivity, electrical infrastructure, and technical support were more limited than in the metropolitan area. In this setting, the project used EVA Classic stations from MakeSens (Sentidos Innovadores e Inteligentes SAS). These low-cost commercial stations were solar-powered and could operate via WiFi when connectivity was available. They measured particulate matter fractions PM1, PM2.5, and PM10, as well as temperature, relative humidity, atmospheric pressure, solar irradiance, and illuminance. Data management was carried out through the MakeSens Cloud platform, which provided a real-time dashboard and access through Python and R interfaces. In this edition, the station was less a construction object and more a pedagogical and observational tool, adapted to a rural environmental education programme (Villarreal-Gómez et al., 2022).

RACiMo-Móncora used EVA Lite stations, also produced by MakeSens, for a metropolitan monitoring network in Bucaramanga. EVA Lite was a compact, portable, solar-powered station optimised for particulate-matter monitoring, especially PM2.5. Its GSM configuration enabled data transmission in locations without WiFi coverage, increasing the flexibility of station siting. The measured variables were similar to those of EVA Classic, and records were transmitted to the MakeSens Cloud platform. In this edition, the hardware was integrated with an online training platform, enabling students to access real network data and analyse it with Python and Jupyter notebooks without requiring high-performance computers (Martínez-Méndez et al., 2022).

In RACiMo-Orquídeas, the project adopted professional commercial equipment to support continuous, low-maintenance, multi-site monitoring across Santander. The network includes Davis Instruments weather stations - Vantage Vue and Vantage Pro2 - and Airlink air-quality sensors. Weather stations record temperature, precipitation, relative humidity, atmospheric pressure, wind direction, and wind speed. Airlink sensors record PM1, PM2.5, PM10, temperature, and relative humidity. Measurements are taken at 15-minute intervals. Station metadata and observational data are accessed via the WeatherLink v2 API and made available through open data repositories and interactive visualisation tools. This edition prioritised robustness, continuity, and regional coverage while maintaining citizen participation through station visits, data access activities, and student-led analysis projects.

## 2.2 Operational model

The operational model of RACiMo can be described as a data chain that begins at the station and ends with community interpretation. At the station level, sensors record environmental variables at predefined sampling intervals. In the early UVA architecture, the Arduino board acquired and pre-processed the signals, while the Raspberry Pi added time information, station metadata, and additional computed quantities such as cloudiness. The system stored records locally and uploaded them periodically. This architecture was designed to make the station autonomous and fault-tolerant, using automatic checks of data recording, connectivity, and upload processes (Peña-Rodríguez et al., 2022).

At the network level, the project progressively moved from local and semi-manual data access toward cloud-based and API-based operation. The prototype and early UVA systems were valuable for training students in electronics, sensor logic, and data acquisition, but they required considerable technical support. The MakeSens stages reduced this operational burden by transferring routine data management to a cloud platform. The Davis Instruments stage further prioritised stability, standardised access, and long-term continuity through WeatherLink-based data acquisition.

At the educational level, each technological stage changed the way students interacted with the monitoring process. In the first editions, the station itself was the central learning object: students learned how sensors, microcontrollers, single-board computers, and repositories were connected. In the most recent editions, the station became part of a broader data infrastructure, and the learning process centred on formulating questions, accessing datasets, producing graphs, interpreting environmental patterns, and communicating results. This shift reflects one of the central design tensions of RACiMo: the more open and constructible the station is, the richer the hardware-learning experience becomes, but the more difficult it is to guarantee long-term stability; the more professional and robust the equipment is, the easier it becomes to sustain a regional monitoring network, but the less direct the student participation in hardware construction becomes.

Thus, the RACiMo equipment history should not be interpreted as a linear replacement of 'inferior' by 'superior' technologies. Each station type responded to a specific educational and territorial context. The DIWO prototype was appropriate for introducing students to open hardware and sensor construction. The UVA stations consolidated an in-house technological platform for local environmental monitoring. EVA Classic and EVA Lite enabled the project to operate in both rural and metropolitan contexts with less reliance on local hardware maintenance. The Davis Instruments network supports the current objective of continuous, multi-municipality monitoring and open access to environmental data. Together, these stages define RACiMo as both a technological and pedagogical infrastructure for citizen environmental monitoring.

# 3. Results and discussion

## 3.1. Evolution of RACiMo: training model, station network, and scale

The evolution of RACiMo can be understood as the progressive transformation of a school-based environmental awareness activity into a broader citizen science programme centred on environmental data. Before RACiMo-Orquídeas, this trajectory included five main stages: the Astroclima precursor, the first RACiMo implementation, RACiMo-Aire, Astropáramo, and RACiMo-Móncora (Table 1). Across these stages, and using the conservative values currently available, the project trained or reached at least 435 students, involved teachers and university mentors, and expanded from urban schools in Bucaramanga to rural communities in the Santurbán páramo and later to a digitally supported metropolitan monitoring network (Asorey et al., 2017a; Asorey et al., 2017b; Peña-Rodríguez et al., 2022; Villarreal-Gómez et al., 2022; Martínez-Méndez et al., 2022)..

The pedagogical model did not remain fixed. It evolved from climate-awareness activities inspired by astronomy and planetary habitability, to hands-on training in open hardware and environmental sensing, then to air-quality monitoring, rural climate education, and finally data-centred learning through online platforms and real network datasets. This evolution reflects a central feature of RACiMo: the stations were never conceived only as instruments for measurement, but as mediating objects through which students could learn how environmental data are produced, organised, analysed, interpreted, and communicated.

### 3.1.1 From climate awareness to citizen environmental monitoring

The initial conceptual basis of RACiMo emerged from Astroclima, an introductory training activity developed between 2013 and 2014. Astroclima connected environmental awareness with astronomical themes such as planetary habitability, exoplanets, Mars exploration, and the greenhouse effect. Its objective was to help students and teachers reflect on the environmental conditions that make life on Earth possible and on the consequences of climate change for those conditions. The course included five sessions, combining conceptual discussions with activities on carbon and water footprints, as well as a hands-on workshop. Approximately 30 students participated in this precursor stage (Asorey et al., 2017b).

Astroclima established the educational premise that later became central to RACiMo: climate change and environmental monitoring can be introduced effectively when global scientific questions are connected to local experience. However, Astroclima remained primarily an awareness-raising and conceptual training activity. RACiMo transformed that approach into a citizen-science model by adding systematic data production, station-based monitoring, open-data practices, and training in data analysis (Asorey et al., 2017a; Peña-Rodríguez et al., 2022).

### 3.1.2 The first RACiMo edition: learning by building and measuring

The first RACiMo implementation, developed in 2015, introduced the project's core methodology: students and teachers worked in teams, accompanied by university mentors, to build and operate monitoring stations and analyse the resulting data. This edition trained approximately 25 students from five schools in the Bucaramanga metropolitan area. The training programme lasted 24 hours and was organised as twelve two-hour sessions (Asorey et al., 2017a; Peña-Rodríguez et al., 2022).

The contents of the first RACiMo training programme reflected the project's dual objectives: scientific literacy and data capacity. The syllabus included six blocks: the information society and collective intelligence; basic statistics and data visualisation; Linux basics and Python programming; environmental variables and sensors; basic electronics and Arduino configuration; and weather-station configuration. The methodology followed a Do-It-With-Others strategy.

Students not only receive information about climate variables; they also participate in configuring the devices, learn the logic of sensor-based measurement, and are introduced to the computational tools needed to inspect and analyse the records (Peña-Rodríguez et al., 2022).

At this stage, the station network was small but pedagogically dense. Its value lay not only in the number of devices installed but also in the fact that each station served as a local learning node. Students could relate measurements to their own school environment, compare records across sites, and understand the difference between isolated observations and a distributed monitoring network. This first edition established the student-centred structure that would characterise later versions of RACiMo.

### 3.1.3 RACiMo-Aire: from climate variables to air-quality awareness

RACiMo-Aire, developed in 2018, expanded the project from weather and climate variables toward air-quality monitoring. It involved seven schools in Bucaramanga and Floridablanca and trained approximately 20 students (Figure 2, left panel). The edition maintained a 24-hour training structure but reoriented its content toward interpreting air pollution and its relevance to urban life.

The training curriculum included citizen science, climatic and environmental variables, air-quality monitoring networks, the Colombian Air Quality Index, data analysis, Python programming, and the operation of monitoring stations. This edition, therefore, introduced a more explicit connection between environmental data and public-health concerns. Students were no longer trained only to understand how climate variables are measured; they were also introduced to interpreting particulate matter records and to using standard categories to communicate air quality conditions (Tarazona-Alvarado et al., 2024).

RACiMo-Aire also changed the project's institutional scale. It incorporated collaboration among the university, public entities, and an industry partner. This academy-industry-government structure strengthened the initiative's public relevance and helped situate school-based monitoring within a broader urban environmental context. Pedagogically, this stage marked a transition from learning by constructing a station to learning how environmental data can inform discussion about air quality, mobility, urban activity, and local decision-making.

### 3.1.4 Astropáramo: rural adaptation and remote methodology

Astropáramo extended the RACiMo experience to rural public schools in the Santurbán páramo region between 2019 and 2021. This stage reached more than 250 students and 19 teachers in rural communities associated with the municipalities of Suratá, Vetas, Berlín, Matanza, and Tona. Its development required a major methodological adaptation because the participating communities had limited access to laboratories, computers, and stable internet connectivity. These limitations became more severe during the COVID-19 pandemic, when most activities had to be conducted remotely (Villarreal-Gómez et al., 2022).

The pedagogical strategy of Astropáramo differed from the urban editions. Instead of beginning with electronics or urban air quality, the programme used astronomy, exoplanet discovery, and astrobiology as entry points to discuss Earth's habitability, climate change, and the ecological importance of the páramo ecosystem. The academic structure was organised into four thematic modules: astronomical observation and unknown worlds; planetary atmospheres; atmospheric physical chemistry; and conditions for life on Earth. Concepts related to statistics, data visualisation, and climate change were incorporated across these modules (Villarreal-Gómez et al., 2022).

A central innovation was the adaptation of the climate-change educational materials from the Klimawandel project of Ludwig-Maximilians Universität München. Experimental kits and astronomical suitcases were prepared to enable students to perform activities autonomously at home or at school. Because synchronous online teaching was often impossible, the methodology combined WhatsApp groups, text messaging, web-based interactions, remote tutoring, individual follow-up calls, and radio-based educational content. These communication channels allowed students to report observations, ask questions, discuss difficulties, and remain connected to the project despite severe connectivity constraints (Scorza et al., 2019; Villarreal-Gómez et al., 2022).

In terms of the station network, Astropáramo altered RACiMo's territorial meaning. The project moved from a metropolitan monitoring context to rural communities located in an ecologically strategic high-mountain ecosystem. The stations served as an introduction to local climate observation and as a bridge between the experimental kits and the environmental conditions of the páramo. This stage showed that RACiMo could be adapted to communities where the main educational challenge was not only access to data, but also access to scientific infrastructure, connectivity, and sustained academic accompaniment.

#### 3.1.5 RACiMo-Móncora: data-centred learning and online scalability

RACiMo-Móncora, developed between 2022 and 2023, returned to the Bucaramanga metropolitan area but with a different pedagogical emphasis. It involved eight educational institutions and approximately 110 students (Figure 2, central panel). Unlike the first RACiMo edition, in which hardware construction was central, Móncora placed greater emphasis on using real monitoring data for programming, visualisation, and analysis.

The main methodological innovation was the use of an online training platform hosted on RedCLARA infrastructure. This platform made the educational materials accessible from standard computers without requiring high technical specifications. It included modules on citizen science, climate change, environmental monitoring, data analysis, Python programming, and Jupyter notebooks. The exercises used data from the Móncora monitoring network, allowing students to work with measurements generated in their own urban environment (Martínez-Méndez et al., 2022).

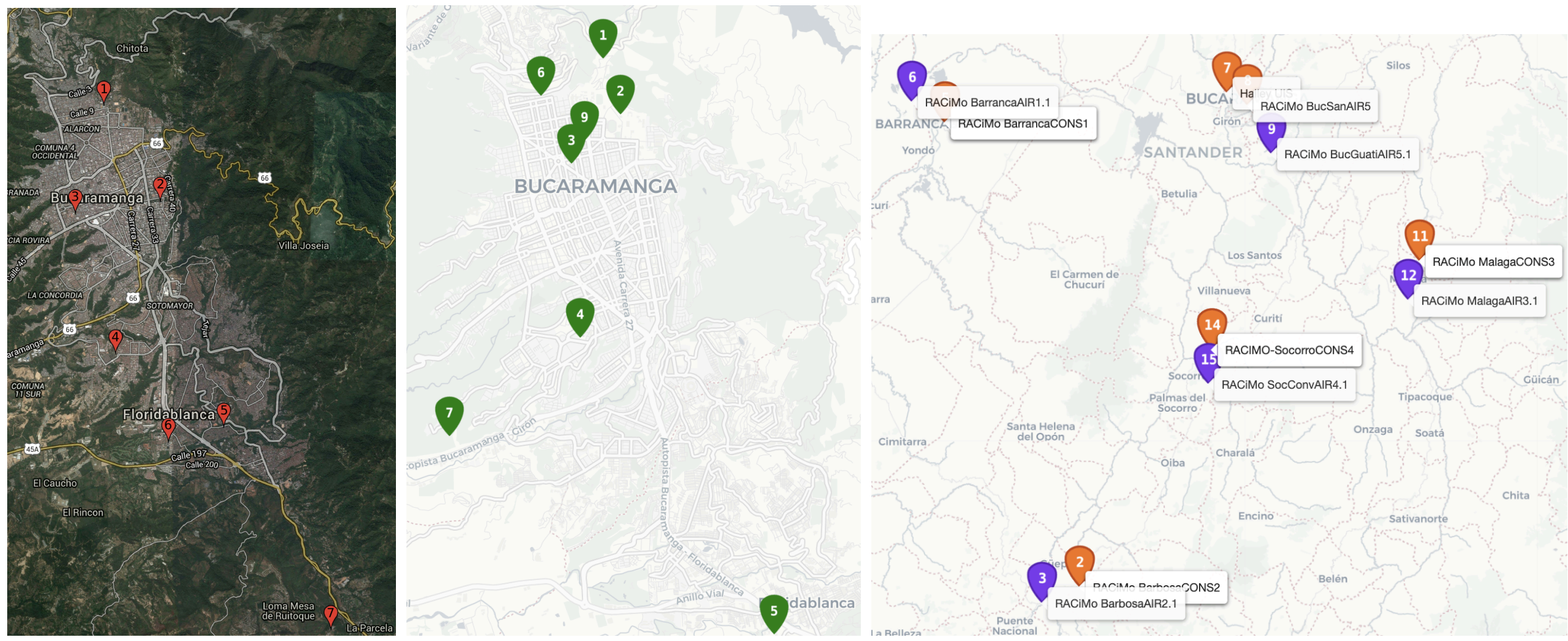

**Figure 2. (Left)** UVA locations at high schools in the metropolitan area of Bucaramanga. *Note.* From top to bottom: (1) Universidad Industrial de Santander, (2) Príncipe San Carlos, (3) Nuestra Señora de Fátima, (4) INEM, (5) Colegio Integrado Andes (CIANDES), (6) New Cambridge and (7) Saucará. (**Centre)** EVA Lite station locations at high schools in the metropolitan zone of Bucaramanga (RACiMo-Móncora). From top to bottom: (1) Instituto Técnico Dámaso Zapata, (2) Colegio Santander, (3) Escuela Normal Superior, (4) Institución Educativa Nuestra Señora del Pilar, (5) Colegio Fundeuis, (6) Institución Educativa Piloto Simón Bolívar, (7) Institución Educativa Luis Carlos Galán Sarmiento, (8) Institución Educativa Café Madrid and (9) Grupo Halley-UIS. (**Right)** Davis Instruments monitoring network across five municipalities of Santander (RACiMo-Orquídeas). Municipalities: Barrancabermeja, Bucaramanga, Málaga, Socorro and Barbosa. Purple markers denote independent Airlink air quality sensors; orange markers denote the WeatherLink consoles (weather stations) coupled with an Airlink sensor.

Móncora also introduced a more explicit project-based structure through research seed groups, or semilleros de investigación, at the participating schools. These groups used station data as a basis for formulating questions, producing graphs, and preparing results for public presentation. The edition culminated in a science fair in 2023, which served as a mechanism for communicating student work and reinforcing the role of schools as active nodes in the monitoring network.

This stage represents a second major pedagogical transition. In the first RACiMo edition, the central learning object was the station as a constructed device. In Móncora, the central learning object became the dataset. Students were trained to access, process, visualise, and interpret environmental records using computational tools. This made the programme more scalable because the training no longer depended primarily on each student's direct interaction with the station's internal hardware.

## 3.2. RACiMo-Orquídeas as the current implementation

RACiMo-Orquídeas represents the current implementation of the RACiMo model and its transition from school-based and metropolitan pilots to a regional citizen-science network. The edition was developed in Santander, Colombia, within the Colombian Ministry of Science, Technology and Innovation's Orquídeas: Mujeres en la Ciencia 2024 call. It was led by a senior female researcher and a young female scientist, and it reached 65% female participation among the students involved. In contrast with earlier editions, where the educational emphasis was strongly connected to building or operating low-cost stations, RACiMo-Orquídeas uses a more robust monitoring infrastructure to support project-based learning, regional comparison, open data, and student-led environmental interpretation.

**Table 1.** Scale and pedagogical emphasis of the RACiMo stages.

| Stage | Period | Students trained or reached | Schools or institutions | Main pedagogical emphasis |
|---|---|---|---|---|
| Astroclima | 2013-2014 | 30 | Not specified | Climate awareness through astronomy, habitability, greenhouse effect, and ecological footprint |
| RACiMo | 2015 | 25 | 5 schools | Citizen science, open data, hardware configuration, sensors, Python, and weather-station operation |
| RACiMo-Aire | 2018 | 20 | 8 schools | Air-quality monitoring, particulate matter, Air Quality Index, data analysis, and urban environmental awareness |
| Astropáramo | 2019-2021 | More than 250 students and 19 teachers | Rural public schools in five municipalities; number of schools to be verified | Climate-change education, astrobiology, experimental kits, remote tutoring, and rural climate observation |
| RACiMo-Móncora | 2022-2023 | 110 | 8 educational institutions | Online training, Python/Jupyter notebooks, real network data, school research seed groups, and science fair |
| RACiMo-Orquídeas | 2025-2026 | 80 | 12 educational institutions | In-person and online training, regional environmental issues, Python/Jupyter notebooks, real network data, school research seed groups, science outreach |

The implementation involved approximately 80 secondary-school and university students from around 17 educational institutions distributed across five municipalities: Barrancabermeja, Bucaramanga, Barbosa, Málaga, and Socorro. These municipalities were selected to broaden the programme's territorial scope and expose students to diverse environmental contexts. Local discussions with researchers, teachers, and community actors helped frame the training around issues such as oil extraction, limestone mining, freight traffic, intensive industrial activity, and regional productive systems. This contextualisation is central to RACiMo-Orquídeas: the objective is not only to collect environmental measurements, but to train students to ask questions about their own territory using real data produced by a shared monitoring network. This follows the original RACiMo principle that citizen science should empower communities to produce, curate, analyse, and use environmental data (Asorey et al., 2017; Peña-Rodríguez et al., 2022).

### 3.2.1 Regional monitoring network

The RACiMo-Orquídeas network was designed to provide continuous meteorological and air-quality measurements across the five participating municipalities (Figure 2, right panel and Suppl. Table 2). The use of professional equipment addresses the need for greater operational stability, lower maintenance requirements, and a network capable of supporting cross-municipal comparisons.

At least one monitoring point in each municipality was located at a Universidad Industrial de Santander site or associated institutional node. When possible, a second station or air-quality sensor was placed in a contrasting setting to capture differences between urban, peri-urban, industrial, or rural conditions. This design allowed students to move from isolated observations to comparative environmental questions. Because low-cost and distributed monitoring systems require careful interpretation and quality control, the project explicitly treated the data as a resource for educational analysis and environmental awareness rather than as a replacement for official regulatory monitoring (Tarazona-Alvarado et al., 2024).

### 3.2.2 Training structure and student project methodology

The training model of RACiMo-Orquídeas was organised in two phases. The first phase provided a common conceptual and computational basis through three five-hour modules: (i) citizen science and the scientific method, (ii) climate change and environmental monitoring, and (iii) descriptive statistics and data analysis with Python. Each module combined one three-hour in-person session with two one-hour online sessions. This hybrid format allowed students to receive direct guidance while also maintaining continuity between in-person activities.

The second phase, devoted to student-led research projects, lasted approximately 15 hours. Students worked in groups, with guidance from junior or senior scientists. Each group selected a question connected to its local environment, designed a basic methodological plan, collected and organised data, produced graphs using computational tools, interpreted the results, and communicated the conclusions in a four-minute science outreach video. The videos were presented in a closing event that brought together the experiences of the participating municipalities.

This project-based approach is the main pedagogical innovation of RACiMo-Orquídeas. The student groups did not simply observe pre-processed figures; they engaged with the data cycle from question formulation to communication. In this sense, the edition consolidated the shift already initiated in RACiMo-Móncora: the station network became the basis for data-centred learning, and the educational objective became the interpretation of local environmental patterns using shared measurements.

#### 3.2.3 Open-data resources and access pathways

Open access to data and educational resources is a defining component of the RACiMo-Orquídeas implementation. The project provides several access levels, from general educational material to real-time station consultation, historical data download, and Python-based data analysis. This tiered structure is important because the intended users have different levels of expertise: school students may begin with visualisation tools, while university students and researchers may access raw or regularly updated datasets for more detailed analyses. The approach is consistent with RACiMo's original open-science orientation and with the MiLAB ecosystem for online learning and data access (Martínez-Méndez et al., 2022).

The current access model also separates real-time observation from reproducible analysis (Table 2). The WeatherLink mobile app associated with the active Davis stations allows users to consult live measurements from active stations, while the Dataverse and open-data repositories support historical analysis and dataset reuse. A public Python script provides a starting point for users who want to access the complete dataset, generate plots, and perform descriptive analyses.

#### 3.2.4 Examples of implementation through student-led analyses

The student projects illustrate how the monitoring network was translated into educational practice (Figure 3). The examples should be interpreted as training exercises in data literacy and environmental reasoning rather than as definitive atmospheric-science studies. Their value lies in showing that students were able to formulate local questions, use real measurements, generate visual representations, and connect the results with territorial conditions.

A first example used wind speed and direction data from Socorro to characterise the local wind regime (Figure 3A). The group produced a wind-rose representation in which the petal length encoded the frequency of occurrence and the radial scale represented mean wind speed. The resulting anisotropy was interpreted in relation to the valley topography surrounding Socorro, which may channel near-surface flow and influence local pollutant dispersion.

A second project explored the relation between ambient temperature and particulate matter in Bucaramanga for PM1, PM2.5, and PM10 (Figure 3B). The three size fractions showed similar behaviour and weak negative correlation with temperature in the analysed records. The students interpreted this as a possible indication of common sources or shared meteorological modulation, although this interpretation should remain tentative without source apportionment, sensor intercomparison, and meteorological controls.

A third project analysed hourly PM2.5 patterns at two Bucaramanga stations (Figure 3C). The students compared the temporal behaviour with air-quality categories derived from Colombian regulatory thresholds, particularly Resolution 2254 of 2017 (Ministerio de Ambiente y Desarrollo Sostenible, 2017). The results suggested higher concentrations during periods of daily mobility, such as morning and evening commuting. This exercise connected data analysis with urban activity patterns and gave students a concrete example of how air-quality measurements may be interpreted in relation to everyday life.

**Table 2.** Open-data and educational resources associated with RACiMo-Orquídeas.

| Resource type | Purpose | Access path |
|---|---|---|
| General educational material | Educational material from the Móncora and Orquídeas editions and contextual information on the monitoring networks. | https://class.redclara.net/halley/moncora/intro.html |
| Interactive data visualisation | Dataverse-based meteorological-data visualiser that allows station selection, parameter configuration, interactive exploration, and download. | RACiMo web page → “Red de Monitoreo RACiMo-Orquídeas” → “Visualizador de Datos Meteorológicos” |
| Real-time data | Live measurements from active stations through the WeatherLink mobile app. | WeatherLink mobile app; users create a free account and add stations by name |
| Historical data | Historical meteorological dataset for 2025–2026. | Datos Abiertos Colombia: https://www.datos.gov.co/; search “RACiMo Orquídeas” |
| Historical data updated daily | Full historical meteorological dataset in CSV format, updated daily. | Dataverse Project, RACiMo collection: https://dataverse.redclara.net/dataset.xhtml?persistentId=doi:10.21348/FK2/UFIOVZ |
| Code for data analysis | Python code to access the dataset, generate plots, and perform basic analyses. | https://n9.cl/sdyra |

A fourth project compared air-quality conditions between stations in Málaga and Bucaramanga using box-and-whisker plots stratified by Air Quality Index categories (Figure 3D). The Málaga stations displayed lower median AQI values concentrated mainly in the “good” category, whereas the Bucaramanga station showed higher medians and broader dispersion. The students related this contrast to differences between a smaller municipality and a larger, more densely trafficked urban environment.

### 3.2.5 Implementation value

RACiMo-Orquídeas demonstrates the project's current maturity in three respects. First, the monitoring network now supports regional comparison across five municipalities rather than isolated school measurements. Second, the educational methodology has moved from station construction toward project-based interpretation of real data. Third, the open-data strategy provides multiple routes for access, from simple visualisation to reproducible computational analysis.

The edition also shows that citizen-science training must be locally contextualised. Although students were generally familiar with terms such as climate change, pollution, and environmental impact, the project revealed that these concepts often required stronger links to everyday experience. The use of municipality-specific environmental issues helped transform abstract environmental discourse into research questions that students could approach with data. This is the central contribution of RACiMo-Orquídeas as a current implementation: it connects robust monitoring infrastructure, open data, and guided student research to strengthen environmental awareness and local capacity for climate adaptation.

## 3.3. Lessons learned and limitations

The successive editions of RACiMo show that a citizen environmental monitoring network is not only a technical infrastructure, but also a social and pedagogical system. Its performance depends on the interaction between the instruments, the data architecture, the training methodology, the continuity of institutional support, and the capacity of participating communities to take ownership of the measurements. From the first open-hardware stations to the current multi-site network, the project has yielded several lessons relevant to future implementations of citizen science in climate and air-quality monitoring.

The main lesson is that the value of RACiMo lies in the full data cycle. The project is strongest when students and teachers are not limited to observing a station, but are guided through the production, curation, analysis, interpretation, and communication of environmental data. This principle was already present in the first RACiMo formulation, which combined station construction, data preservation, and training in electronics, programming, statistics, and environmental interpretation (Asorey et al., 2017a; Peña-Rodríguez et al., 2022). However, the implementation of that principle has required different compromises in each technological and territorial context.

### 3.3.1 Trade-off between hardware participation and network reliability

The first RACiMo editions demonstrated the pedagogical strength of open hardware. The Do-It-With-Others methodology allowed students to understand sensors as physical and electronic systems rather than as opaque measuring devices. By assembling, configuring, and calibrating stations, participants could connect environmental variables to sensor responses, data files, metadata, and computational analyses. This model was particularly effective for introducing students to open data science and for reducing the distance between school communities and university-based research (Asorey et al., 2017a; Peña-Rodríguez et al., 2022).

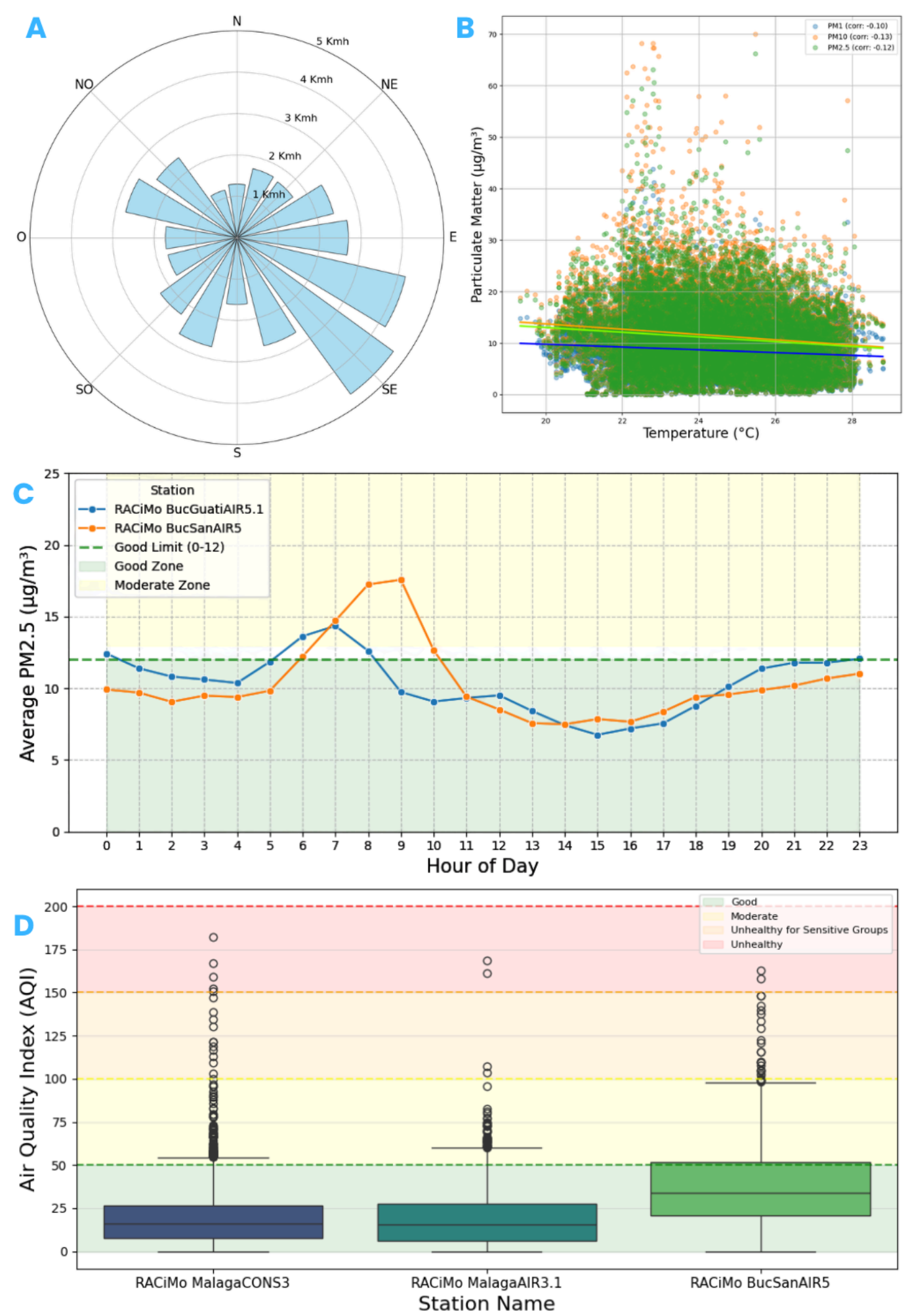

**Figure 3.** Examples from student-driven projects from the RACiMo-Orquídeas edition, representing different municipalities and themes. (A) Wind rose characterising the local wind regime in Socorro, where petal length encodes the frequency of occurrence and the radial scale the mean wind speed (km $h^{-1}$) for each direction. (B) Relationship between particulate matter (PM1, PM2.5 and PM10) and ambient temperature in Bucaramanga. (C) Mean PM2.5 concentration as a function of the hour of the day for two stations in Bucaramanga (RACiMo BucGuatiAIR5.1 and RACiMo BucSanAIR5), with the air-quality thresholds of Resolution 2254 of 2017. (D) Air Quality Index across stations in Málaga (RACiMo MalagaCONS3 and RACiMo MalagaAIR3.1) and Bucaramanga (RACiMo BucSanAIR5), shown as box-and-whisker plots stratified by the standard AQI categories.

The same approach also created operational limitations. Open, modular, low-cost stations require frequent technical supervision, careful calibration, protection against environmental exposure, and sustained maintenance. In a school-based network, this burden cannot be transferred entirely to teachers or students. As the project expanded, the need for continuity, remote access, and reduced maintenance led RACiMo toward commercial and professional instruments. This increased robustness and improved the feasibility of long-term monitoring, but reduced the degree of direct student participation in hardware construction.

This trade-off should be made explicit in the manuscript. Hardware openness and educational depth are not automatically aligned with network stability. For future versions, a hybrid model is preferable: maintain simplified open-hardware modules for training and prototyping, while using more robust stations for continuous public datasets.

### 3.3.2 Data quality, calibration, and comparability

RACiMo's early technical development included an important effort in calibration and commissioning. The UVA stations were tested by comparing them with reference instruments, including temperature with a Fluke 1524 Temperature Meter, humidity between stations under common conditions, irradiance with an Ambient Weather WS-1400-IP station, and cloudiness estimation via image segmentation and comparison with irradiance (Peña-Rodríguez et al., 2022). These tests show that the first stations were not merely educational artefacts; they were operational prototypes whose limitations were experimentally examined.

Nevertheless, data quality remains one of the central limitations of the project. Low-cost environmental sensors are useful for increasing spatial coverage and educational purposes, but their records require calibration, metadata documentation, periodic validation, and careful interpretation, especially when used for air-quality assessment (Kumar et al., 2015; Mead et al., 2013; Tarazona-Alvarado et al., 2024). In RACiMo, the technological evolution from prototype stations to UVA, EVA, and Davis instruments also implies that historical datasets are not automatically homogeneous. Differences in sensors, sampling intervals, siting conditions, calibration procedures, maintenance practices, and data platforms must be documented before long-term comparisons are made.

For this reason, RACiMo data should be presented as a citizen-science and educational monitoring resource, not as a regulatory-grade environmental record unless the corresponding validation has been performed.

### 3.3.3 Failures as methodological information

One of the most important lessons from RACiMo is that failures should be documented as part of the scientific and educational record. In citizen science, interruptions, incomplete data, connectivity problems, installation errors, and maintenance difficulties are not marginal details; they define the conditions under which a monitoring network can operate.

Astropáramo provides the clearest example. The edition successfully adapted climate-change education to rural communities during the COVID-19 pandemic through experimental kits, WhatsApp groups, remote tutoring, and radio-based educational support. However, the associated EVA stations did not generate data that students could use for plotting and analysis due to installation faults, even though students with access to computers received training in their use (Villarreal-Gómez et al., 2022). This limitation is pedagogically and technically relevant. It shows that the delivery of equipment is not equivalent to operational monitoring and that rural implementation requires additional procedures for installation verification, remote diagnostics, local technical support, and fallback activities when stations do not produce usable records.

### 3.3.4 Sustainability and institutional continuity

The continuity of RACiMo over several editions has depended on a succession of funding sources and institutional alliances, including university support, public grants, municipal participation, industry collaboration, and international educational partnerships. This diversity has allowed the project to adapt to changing priorities and contexts. It has also made RACiMo resilient: when one edition emphasised hardware construction, another emphasised air quality, rural climate education, online training, or open data. At the same time, sustainability remains a limitation. Monitoring networks require long-term responsibilities that extend beyond the duration of a grant: station maintenance, sensor replacement, data hosting, repository preservation, platform administration, ethical management of student outputs, and continuity of school relationships.

### 3.3.5 Replicability and conditions for transfer

RACiMo offers a replicable model, but replication should not be interpreted as simply installing the same equipment elsewhere. The transferable component is the integration of monitoring stations, open or accessible data infrastructure, educational materials, and guided student projects. However, the conditions for replication vary strongly across contexts. Urban schools may have better connectivity but greater air-quality complexity; rural schools may have strong environmental relevance but limited internet access, fewer computers, and greater logistical constraints. A successful transfer, therefore, requires a preliminary diagnosis of local infrastructure, teacher availability, connectivity, environmental concerns, safety conditions for station installation, and expected maintenance capacity.

The replicable unit of RACiMo should be described as a methodological package rather than a fixed instrument: a training sequence, a station or data source adapted to the context, a data-access protocol, a set of analysis activities, and a strategy for public communication of student results.

## 4. Conclusions and perspectives

RACiMo has evolved from a school-based open-hardware experiment into a regional citizen-science programme for environmental monitoring, climate awareness, and data literacy. The project has built a pedagogical and technological pathway through which students, teachers, university mentors, and local communities learn to record environmental variables, curate the resulting datasets, analyse them with open computational tools, and interpret them in relation to local climate and air-quality concerns.

RACiMo's evolution reflects a deliberate rebalancing among openness, educational depth, and operational reliability. Early stations immersed students in hands-on hardware construction — sensors, microcontrollers, calibration, and local data analysis — while later stages introduced more robust instruments to sustain broader, continuous monitoring networks. Pedagogically, the project grew from linking climate awareness to astronomy and planetary habitability, through Do-It-With-Others station building and air-quality monitoring, to student-led research grounded in real network data. Across its editions, RACiMo has reached approximately 515 participants — secondary-school and university students, rural communities, teachers, and institutional partners — offering evidence that citizen science functions as a sustained educational strategy when adapted to each community's conditions rather than imposed as a fixed model.

RACiMo-Orquídeas represents this model's current maturity: by deploying a network of weather stations and air-quality sensors across five municipalities in Santander and publishing data through open repositories and visualisation tools, it consolidates the shift from isolated school stations to a department-scale environmental data infrastructure whose student projects continue to demonstrate its educational value.

The limitations of RACiMo are also part of its contribution. Low-cost and citizen-operated sensors aiming to produce research-quality data require calibration, metadata, quality-control procedures, maintenance logs, and cautious interpretation, especially when air-quality data are involved. We also highlight that equipment delivery does not automatically produce usable data: connectivity constraints, installation failures, and limited local technical support can interrupt the monitoring chain even when the educational intervention remains valuable. To further evaluate RACiMo's educational impact, incorporating the analysis of student surveys and longitudinal follow-up of participants will be key to better understand learning outcomes, changes in environmental understanding, and possible effects on scientific identity.

From a public-policy perspective, RACiMo suggests that regional and municipal governments can strengthen climate-awareness strategies by supporting citizen environmental monitoring as part of local education and adaptation programmes. Such support should not be limited to purchasing sensors. It should include teacher training, school-based data-literacy activities, open-data infrastructure, maintenance budgets, technical support, and mechanisms for connecting citizen-generated observations with official environmental information systems. When properly framed, citizen monitoring can help communities understand the environmental conditions of their own territories, identify locally relevant questions, and participate more actively in discussions about climate adaptation, air quality, and territorial planning.

## Acknowledgments

The authors express their special gratitude to Jorge H. Martínez Téllez for his sustained support of RACiMo and for his continued efforts to secure funding for the development and consolidation of the project. His commitment has significantly strengthened the School of Physics and its outreach, education, and citizen-science initiatives.

We thank the Colombian Ministry of Science, Technology and Innovation for supporting RACiMo-Orquídeas through the Orquídeas: Women in Science 2024 call 948 (*Convocatoria Orquídeas: Mujeres en la Ciencia 2024*), under project code No. 109309. We also gratefully acknowledge the Regional Fund for Digital Innovation in Latin America and the Caribbean (FRIDA), which supported the first edition of RACiMo under project 347.

The RACiMo-Aire edition was supported by Colciencias and the Colombian Ministry of Information and Communications Technologies. We also thank Multiprocesos SIG S.A.S., the Bucaramanga and Floridablanca town halls, and the corresponding municipal partners for their collaboration. Astropáramo was financed by the Acueducto Metropolitano de Bucaramanga (AMB) and the Cámara de Comercio de Bucaramanga through the *Investigadores por Naturaleza* institutional programme.

We acknowledge the Universidad Industrial de Santander for its support through the Office of Research and Extension, the Institute for Regional Projection and Distance Education (IPRED), the Faculty of Sciences, and the School of Physics, which contributed to RACiMo-Móncora and other editions of the project. Finally, we thank all teachers, students, university mentors, and school communities who have participated in RACiMo since 2013. Their enthusiasm and dedication have been essential to the continuity and impact of the project.

## Author biographies

**Laura V. Flórez** is a postdoctoral researcher at the Universidad Industrial de Santander and leads the RACiMo-Orquídeas project. She holds a doctoral degree from Friedrich Schiller University Jena and the Max Planck Institute for Chemical Ecology in Germany. Her research experience includes postdoctoral work at the University of Copenhagen and project leadership at Johannes Gutenberg University Mainz. Her work connects plant and environmental sciences, evolutionary ecology, science education, and citizen-science initiatives. She is also an active member of ScienteLab, a Colombian scientific diaspora organisation that promotes science education among children and young people.

**Gabriela Sánchez-Ariza** is a physicist and Young Researcher in the RACiMo project, supported by Colombia's Ministry of Science, Technology and Innovation. Her research interests include environmental data analysis, computational astrophysics, and tidal disruption events. She has conducted research at the National Institute for Astrophysics, Optics and Electronics (INAOE, Mexico) and has presented her work on tidal disruption simulations at international scientific conferences.

**Nicolas Mantilla-Molina** is a physics researcher who completed his undergraduate studies in Physics at the Universidad Industrial de Santander (UIS) in Bucaramanga, Colombia. He is an active member of the Halley group, where he specialises in theoretical and computational physics, focusing on Relativistic Mean Field theory and sigma-omega models. Concurrently, he served as a scientific counsellor for regional citizen science initiatives, where he managed data operations and led the data analysis for local environmental monitoring station networks. He is currently conducting research at the TRIUMF laboratory in Vancouver, Canada, focusing on In-Medium Similarity Renormalization Group methods

**Jhonattan J. Pisco-Guabave** holds a degree in Mathematics and Physics from the University of Los Llanos and in Mechanical Engineering from the Universidad Industrial de Santander, where he currently works as a faculty member in the School of Physics. His work combines research, outreach, and education in astronomy, physics, and citizen science. He has played a leading role in RACiMo, helping connect secondary-school students in Bucaramanga with open-source tools for environmental monitoring, data analysis, and science education.

**Alexander Martínez-Méndez** is a doctoral candidate in Computer Science at the Universidad Industrial de Santander. He serves as the Software Architect and Systems Administrator for the EL-BONGÓ Physics project, where he manages the ElBongo Science Gateway, which supports open science and reproducible research across Latin America. His expertise centres on the Computing Continuum, workflow orchestration, and scientific data management. With extensive experience designing distributed research platforms, including RedCLARA's MiLab, he has co-authored publications focusing on high-performance computing, data integrity, and computational reproducibility.

**Alexandra Serrano-Mendoza** is an undergraduate Physics student at the Universidad Industrial de Santander and a member of Grupo Halley and the Red de Estudiantes Colombianos de Astronomía. Since 2021, she has participated in science outreach, education, and research projects. Within RACiMo, she contributed to AstroPáramo by organising educational and digital resources and later supported the Móncora project through curriculum design, science clubs, and student-led activities. She has also worked on environmental and astronomy research, including studies of particulate matter and satellite streak identification.

**Luis A. Núñez** is Professor Emeritus at the Universidad de Los Andes in Mérida, Venezuela, and Full Professor in the School of Physics at the Universidad Industrial de Santander, Colombia. His research interests include relativistic astrophysics, astroparticle physics, and information sciences. He has co-authored nearly 350 articles in international journals. He is the principal investigator and spokesperson of the Latin American Giant Observatory and Colombia's representative in the Pierre Auger Observatory Collaboration, within the CosmoGeophysics team. He is also a co-principal investigator of ERASMUS+ projects in physics higher education.

**Hernán Asorey** is a physicist, deep-tech entrepreneur, and science-to-market builder. He received his PhD from Instituto Balseiro, Centro Atómico Bariloche, in 2012. His work spans radiation detection, astroparticle physics, muography, space weather, medical physics, scientific computing, IoT, blockchain, and data science. He was formerly Principal Investigator A at CNEA and Head of its Medical Physics Department. He has authored more than 230 peer-reviewed publications, an introductory physics textbook, and a patent related to water Cherenkov detectors. He is currently founder of piensas.xyz and Fractalis Adventures.

**Laura M. Becerra Bayona** is a relativistic astrophysicist and Assistant Professor at Universidad Mayor de Chile. Her research explores compact objects, high-energy astrophysical phenomena, magnetic fields, accretion processes, and numerical simulations. She earned her PhD in Physics and Relativistic Astrophysics from Sapienza Università di Roma in 2018. Her doctoral thesis on hypercritical accretion in compact stars was recognised with the 2018 International Astronomical Union PhD Prize in High Energy Phenomena and Fundamental Physics, one of the IAU's distinctions for outstanding doctoral research in astronomy.

**Jennifer Grisales-Casadiegos** is a PhD candidate in Astrophysics at the Universidad de Guanajuato, Mexico. She received her BSc and MSc in Physics from the Universidad Industrial de Santander, Colombia, where she worked on astroparticle physics within the LAGO and Pierre Auger collaborations. With the Halley Astronomy Group, she has developed projects in science education and social appropriation of knowledge. She led AstroPáramo, an educational initiative adapted from a climate-change curriculum developed at Ludwig-Maximilians-Universität München, and integrated its weather-station component into the RACiMo citizen-science network.

**Arturo Núñez-Castiñeyra** is a researcher at the Institut d'Astrophysique de Paris. His work combines cosmological simulations, galaxy-formation theory, dark-matter phenomenology, cosmic-ray physics, and high-performance computing. He received his PhD from Aix-Marseille University, where he studied the connection between cosmological simulations and dark-matter detection. He later joined CEA Saclay, working on cosmic-ray transport, the interstellar medium, and the impact of turbulence on stellar populations. His current research focuses on r-process enrichment and binary neutron-star mergers in cosmological simulations.

**Jesús Peña-Rodríguez** is an electronics engineer with a master's degree in Electronic Engineering and a PhD in Physics. He is currently a researcher at the Faculty of Physics, University of Wuppertal, Germany, where he contributes to the Compressed Baryonic Matter Experiment. He is also involved in the ALICE Collaboration through the ALICE 3 project. His research interests include particle detector development, electronic instrumentation, radiation detection systems, data acquisition, and experimental nuclear and particle physics. He is a member of the DRD4 Collaboration at CERN and participates in the NRW-FAIR Network.

**José Salamanca Coy** graduated from the Universidad Industrial de Santander, Colombia. His career began working directly with the RACiMo project, which inspired the founding of MakeSens as a spin-off — designing low-cost monitoring stations (EVA, EVA Light, ERIS, CAMI) deployed across RACiMo projects in Bucaramanga and Latin America. He continues to drive open, community-based smart environmental monitoring initiatives, and also serves as Lead AI Engineer at Source Meridian.

**Christian Sarmiento-Cano** is a professor at the Universidad Autónoma de Bucaramanga and an adjunct professor at the Universidad Industrial de Santander. His research focuses on astroparticle physics, cosmic rays, and their applications to geophysics and agriculture. He is Colombia's representative in the LAGO collaboration and a former member of the Pierre Auger Observatory. He has authored more than 50 scientific publications and has experience mentoring students and early-career researchers in astroparticle physics, environmental applications, and collaborative scientific projects.

**David Sierra-Porta** is an Associate Professor at Universidad Tecnológica de Bolívar, Cartagena de Indias, Colombia, where he is affiliated with the School of Digital Transformation and the Department of Basic Sciences. He directs undergraduate and graduate programmes in Data Science, Applied Statistics, and Applied Statistics and Data Science. He also leads the Gravitation and Applied Mathematics research group. His research focuses on time-series analysis, complex networks, graph theory, machine learning, atmospheric physics, cosmic rays, space weather, and solar dynamics.

**Supplementary Material**

**Supplementary Table 1**. Summary of monitoring devices and their evolution throughout the history of RACiMo.

| RACiMo stage | Main equipment | Operational logic | Main variables | Data access and transmission |
|---|---|---|---|---|
| RACiMo prototype | Arduino Uno, Raspberry Pi, GPS, WiFi, sensor shield | DIWO construction with students; low-cost modular station | Temperature, pressure, illuminance | WiFi transmission to central repository; local analysis with Python |
| UVA / RACiMo-Aire | In-house UVA station | Modular autonomous station built by research team; improved reliability | Pressure, temperature, humidity, precipitation, irradiance, illuminance, cloudiness; later PM2.5 and PM10 | Local storage, metadata files, web upload, exportable datasets |
| Astropáramo | EVA Classic, MakeSens | Solar-powered station for rural schools; reduced hardware interaction | PM1, PM2.5, PM10, temperature, humidity, pressure, irradiance, illuminance | MakeSens Cloud dashboard; Python/R API |
| RACiMo-Móncora | EVA Lite, MakeSens | Compact solar station; GSM communication for sites without WiFi | PM2.5-focused air-quality monitoring plus meteorological variables | MakeSens Cloud; online training platform; Jupyter/Python exercises |
| RACiMo-Orquídeas | Davis Vantage Vue, Vantage Pro2, Airlink 7210 | Professional low-maintenance network for multi-municipality monitoring | Temperature, precipitation, humidity, pressure, wind speed, wind direction, PM1, PM2.5, PM10 | WeatherLink v2 API; open-data repository; interactive visualisation tool |

**Supplementary Table 2.** Monitoring nodes in the RACiMo-Orquídeas network.

| Station name | Main device(s) | Station ID | Municipality | Role in the network |
|---|---|---|---|---|
| Halley UIS | Vantage Pro2 | 84759 | Bucaramanga | Meteorological reference station at UIS |
| RACiMo BarrancaCONS1 | WeatherLink console, Vantage Vue 6242, Airlink 7210 | 219664 | Barrancabermeja | Weather and air-quality node |
| RACiMo BarbosaCONS2 | WeatherLink console, Vantage Vue 6242, Airlink 7210 | 219666 | Barbosa | Weather and air-quality node |
| RACiMo MalagaCONS3 | WeatherLink console, Vantage Vue 6242, Airlink 7210 | 219667 | Málaga | Weather and air-quality node |
| RACiMo SocorroCONS4 | WeatherLink console, Vantage Vue 6242, Airlink 7210 | 219668 | Socorro | Weather and air-quality node |
| RACiMo BucSanAIR5 | Airlink 7210 | 219677 | Bucaramanga | Independent air-quality node |
| RACiMo BucGuatiAIR5.1 | Airlink 7210 | 219678 | Piedecuesta/Bucaramanga | Independent air-quality node |
| RACiMo BarrancaAIR1.1 | Airlink 7210 | 219679 | Barrancabermeja | Independent air-quality node |
| RACiMo BarbosaAIR2.1 | Airlink 7210 | 219680 | Barbosa | Independent air-quality node |
| RACiMo SocConvAIR4.1 | Airlink 7210 | 219681 | Socorro | Independent air-quality node |
| RACiMo MalagaAIR3.1 | Airlink 7210 | 219682 | Málaga | Independent air-quality node |